\documentclass[referee]{aa} 
\usepackage{graphicx}
\usepackage{txfonts}
\begin{document}
   \title{Detection of mesogranulation at the upper chromosphere from SOHO/SUMER observations}

   \author{{R. Kariyappa\inst{1}, B.A. Varghese\inst{1}} 
          \and
          W. Curdt\inst{2}}
   \institute{Indian Institute of Astrophysics, Bangalore 560 034, India\\
              \email{rkari@iiap.res.in}
         \and
             Max-Planck-Institut f\"ur Aeronomie, D-37191 Katlenburg-Lindau, Germany\\}
\offprints{R. Kariyappa}

   \date{Received ........................ / Accepted ....................... }

 
  \abstract
{}  
   { Our aim is to detect the mesogranulation structures at the upper solar chromosphere}
   {We have analyzed a time series of spectra in the hydrogen Lyman lines and the
Lyman continuum obtained
by the Solar Ultraviolet Measurements of Emitted Radiation (SUMER) spectrometer
on the SOlar Heliospheric Observatory (SOHO).  The
time series of about 2 hours and 22 minutes was obtained on 1999 March 9 in a quiet
region near the center of the solar disk.  For our analysis, we have selected a
Lyman continuum window around 907 \AA, and the five Lyman lines: Ly5 (937.8 \AA),
Ly7 (926.2 \AA), Ly9 (920.9 \AA),
Ly11 (918.1 \AA), and Ly15 (915.3 \AA).
 A Fast Fourier analysis has been performed
in the spatial domain, all along the slit length used, for all the spectra and
for
the total duration of the observations.}
   {We have detected a significant periodic
spatial variations with Fourier transform power peaks around 9-10 arcsec and at 4
arcsec.  They correspond to the scale of the mesogranulation structure and the
width of the supergranular boundary,
respectively.   For the first time, this provides evidence
for the existence of a meso-scale in the upper chromosphere, of the same size
as observed in the photosphere and lower chromosphere by earlier studies.
 We find from the analysis
that there seems to be no signature of any temporal evolution associated with the
mesogranules, at least not
during our observing period. This result suggests that the life time of
mesogranules will be several hours or more, which confirms the earlier
findings.  In addition, we notice that the
size (9-10 arcsec)
of the mesocells
appears to be the same in all Lyman lines and in the
continuum, which are formed at different depths in the chromosphere.} 
   {}


\titlerunning {Detection of mesogranulation in upper chromosphere}
\authorrunning {R. Kariyappa, et al.}
   \maketitle
    
%

\section{Introduction}

It is well known that there is evidence for at least three scales of motion in the solar
photosphere: granulation, mesogranulation, and supergranulation.  The solar surface is subject
to convective and oscillatory motions, the properties of which are fundamental
to understanding the physics of solar surface evolution.  One of the striking
properties of the observed convective motions on the Sun is the discrete
spectrum of their horizontal scales.  Granules have scales of order 1 arcsec,
the mesogranules of order 8 arcsec, and
supergranules of order 40 arcsec.
 Granulation has been the
subject of studies for more than a century, and its structure, physical
properties, the life time, and evolution are rather well known both from observational work
and numerical models, except at the smallest spatial scales below 0.2",
which are at present unaccessible by both ground-based and
space-borne solar telescopes.

Simon \& Leighton (1964) have discovered the supergranulation as a
large-scale convective
pattern, but the question of just what determines
its dominant horizontal scale remains open.
 Muller et al. (1992) have studied the evolution of solar mesogranules,
which seem to survive for at least 3 hours.
In a recent paper, Roudier et al. (1998) find
the lifetimes for mesogranules will be ranging from 16 to 185 min.
  November et al. (1981) suggested that mesogranulation
could come from He$^{0}$  ionization at a depth of 7 Mm, as the "missing"
scale in the scenario of the solar convective motions
(in between the
granulation arising from the H$^{0}$ ionization instability region at 2 Mm
depth, and the supergranulation from the ionization of He$^{+}$ at 20 Mm).
 From the studies using simultaneous
observations of the granular flows (SOUP instrument on board SPACELAB 2)
and the magnetic field (Big Bear Observatory), Simon et al. (1988) raise the
question "whether the mesogranule is really a scale of solar convection
distinct from the supergranulation".
 There is still
disagreement on the nature of the mesogranulation: whether it represents waves
(Dam\'{e} 1985, Straus \& Bonaccini, 1997) or convection
(e.g., Deubner 1989; Straus, Deubner, \& Fleck 1992);
whether it is a distinct spatial regime (Title et al. 1986; Simon et al. 1988,
Ginet and Simon 1992) or cannot be identified as regime separated from granulation
(Straus et al. 1992, Straus \& Bonaccini 1997).  Therefore, all
these issues need to be better understood and related to the observed
properties of the mesogranulation, particularly from a long-time sequence of
simultaneous observations in many spectral lines.

         The overshooting of convective motions into the solar
photosphere, chromosphere, and transition region may provide a direct coupling
between the atmosphere and the vigorous turbulence below the surface.  Following
the observation and a first attempt of interpretation of the mesogranulation by
November et al. (1981), several related observations were carried out using different
techniques either through intensity measurements (Koutchmy \& Lebecq 1986), statistical
studies of active granules grouping (Oda 1984), or by measurements of granular
flows (November et al. 1987).
 Recently, Hirzberger et al. (1999) have
studied a time series of white-light images of 80-minute duration and have shown
that the dynamics of exploding granules is strongly affected by their
surroundings and that their appearance is closely related to the mesogranular
flow field.  Martic, Dam\'{e}, \& Kariyappa (2000, in preparation) have analyzed the solar ultraviolet filtergrams
obtained during the fourth rocket flight of the Transition Region Camera (TRC) of
1985 October 25,
 and have derived a mesogranulation pattern with a size of 8 Mm in the temperature
minimum region.  From different observations in \ion[Ca ii] H \& K lines
(Dam\'{e} \& Martic 1987,
 1988; Kariyappa, Sivaraman, \& Anadaram  1994) the meso-scale phenomenon was identified
in the lower chromosphere.

In this paper, we attempt to detect the meso-scale structure
in the upper chromosphere using a long-time sequence of spectra
obtained with the SOHO/SUMER spectrometer in five Lyman lines and the continuum.
They have simultaneously been observed near the center of the solar disk in a quiet region.


\section{Observations and methods of data analysis}

  
SUMER is a stigmatic, high-resolution normal-incidence spectrometer
that covers the wavelength range between 400 and 1610~\AA.  With a spatial
resolution close to 1 arcsec and spectral pixels of $\approx$43~m\AA (with
subpixel
resolution), it provides a unique opportunity to observe the whole
hydrogen Lyman series plus the Lyman continuum.  The SUMER capabilities are
described in detail by Wilhelm et al. (1995), the in-flight performance is
presented in Wilhelm et al. (1997) and Lemaire et al. (1997), and specific
aspects related to the Lyman series spectra by Curdt \& Heinzel (1998), Heinzel \& Curdt
(1999), Curdt et al. (1999) and Kariyappa, et al. (2001).

        In this paper, we have analyzed the data set observed on 1999 March 9.  We
used the 1" $\times$ 120" slit. We have obtained a set of 306 spectra with a
repetition rate of about 27.72 s.  A few extra seconds were required for the
solar rotation tracking, using a special, so-called half-step solar rotation
compensation scheme (cf., Curdt \& Heinzel 1998). The whole sequence thus lasted for
8480~s ($\approx$ 2 hours and 22 minutes).
The slit has crossed two network cells with less brightness, a very bright network
boundary lane and finally ends with a weak network lane on the southern part
of the slit.  In between these network regions, there are
internetwork regions with oscillatory nature.  The spectra were acquired by the detector A on its
central band of 120 spatial x 1024 spectral pixels.  This covered the
wavelength band from 903 to 943 \AA, including all Lyman lines from
Ly5 up to the series limit,
plus a portion of the Lyman continuum.  This is shown in Figure 2 of Curdt and Heinzel (1998).
 We have selected five hydrogen Lyman lines:
Ly5 (937.8 \AA),
Ly7 (926.2 \AA), Ly9 (920.9 \AA),
Ly11 (918.1 \AA) and Ly15 (915.3 \AA), and a
Lyman continuum window around 907 \AA \ for the detailed study.  For detailed
line identifications see Curdt et al. (1997) and Curdt \& Heinzel (1998).

        The data set has been decompressed, flat-field corrected, and
corrected for the geometrical distortion of the detector.  We have performed
a spatial Fourier analysis  all along the slit length (120 arcsec)
for all the Lyman lines and the continuum, and at every time interval for the
duration of our observations to determine the periodicity in the spatial scale.


\section{Results and discussions}

\subsection{Oscillatory behavior in bright network and internetwork regions}

It is evident from the data set of 1999 March 9 that there
are significant temporal variations of the intensities and shapes of all
Lyman lines observed.
We have derived the central intensity values for all the line profiles
after calculating the center of gravity of the average profiles of these
lines and applying a smooth--3 noise filter using all 306 spectra.

 \begin{figure}
   \centering
  \includegraphics[width=13.0cm,height=9.5cm]{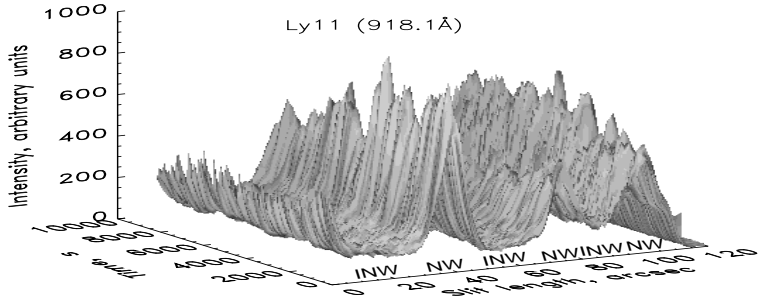}
   \caption{Ly11 (918.1 \AA) central intensity map averaged by a
smooth-3 noise filter and shown as three-dimensional plot.  Here we have marked
the network and internetwork regions as NW and INW respectively.  We
notice the intensity oscillatory nature in INW and NW.}
\label{FigVibStab}
    \end{figure}

\begin{figure}
   \centering
  \includegraphics[width=13.0cm,height=9.5cm]{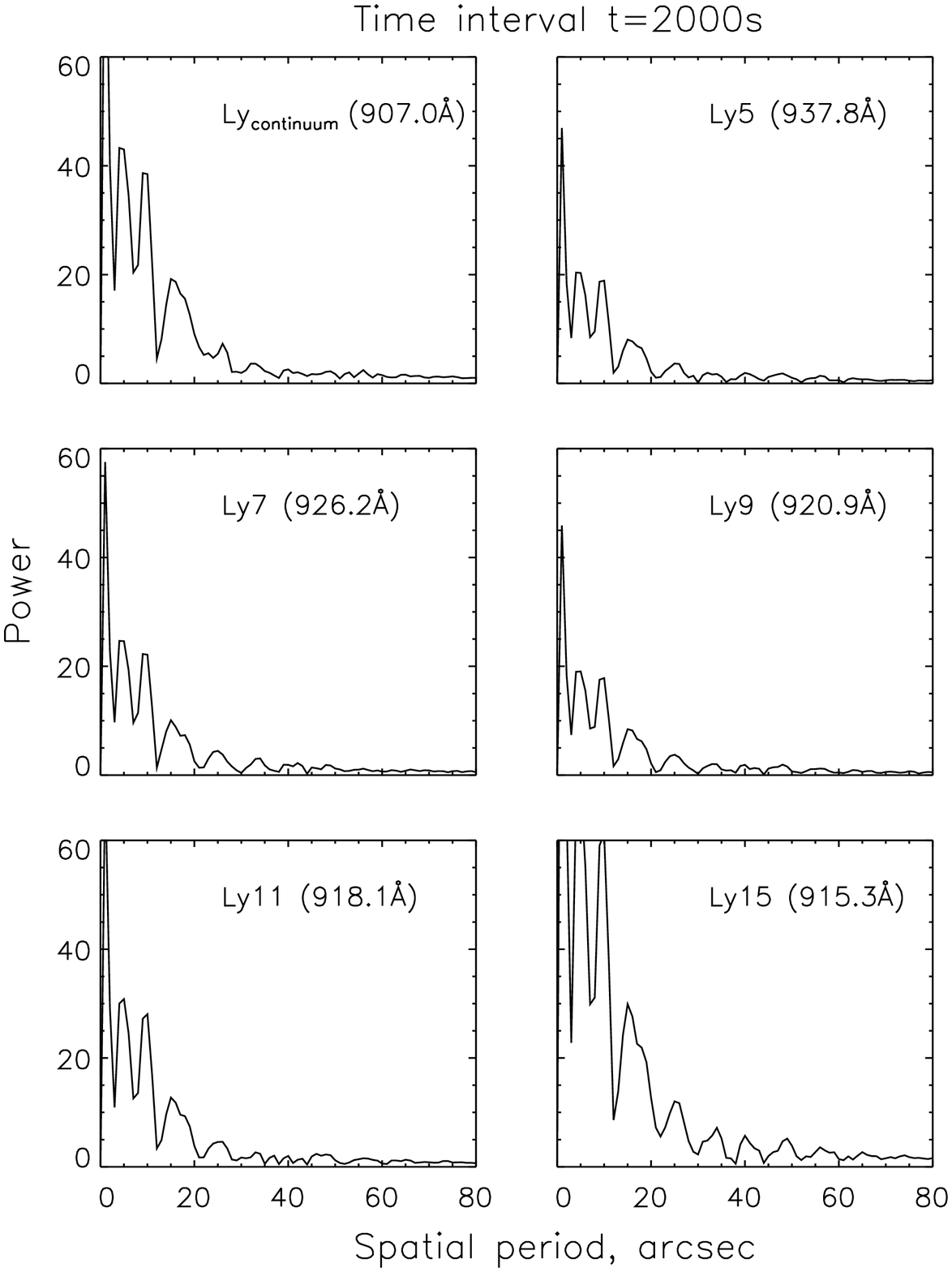}
   \caption{Spatial power spectra of Ly continuum (907.0 \AA),
Ly5 (937.8 \AA), Ly7 (926.2 \AA), Ly9 (920.9 \AA), Ly11 (918.1 \AA),
and Ly15 (915.3 \AA) for the time interval around 2000 s during the sequence.
Note that there are two significant peaks around 9-10 arcsec (meso-scale)
 and at 4 arcsec (width of the supergranular boundary).
Also notice that these two peaks are well dominant in all the lines, which are formed
at different depths in the upper chromosphere.}
\label{FigVibStab}
    \end{figure}

       In Figure 1, we have shown a 3-D plot of the central intensity versus
time and distance along the slit for the Ly11 (918.1 \AA) line.
It can be seen that the slit crosses three network
lanes, which we have marked as NW, one of them is very bright at a
pixel position around 40 and
other two are less bright (at pixel positions roughly around 72 and 97).
In between the network lanes, we see features of the internetwork
oscillatory grains (around pixel positions : 10-20, 50-60, and 84),
which we marked as INW.  We can see
 from the 3D-plot that internetwork (INW) regions will show an oscillatory
behavior of periodicity around 3-min, which one can derive by counting the number of intensity
peaks over the observed time period.  In a similar way,
the network regions show 5-7 min. periodicity in their intensity oscillations.
The 3-min in internetwork and 7-min in network regions have been reported earlier by
Curdt \& Heinzel (1998) and Kariyappa et al. (2001) from an analysis of a long time series spectra
of hydrogen Lyman lines obtained with the
SOHO/SUMER experiment.  A large number of people have reported similar
results from \ion[Ca ii] H \& K lines ( e.g., Liu 1974; Cram \& Dam\'{e} 1983;
Lites, Rutten, \& Kalkofen
1993; Kariyappa et al. 1994; Kariyappa 1994, 1996).  It is important to note that the emission profiles
of the optically thick Lyman lines span relatively large atmospheric depths
(see, e.g., Fig.1 in Vernazza, Avrett, \& Loeser 1981) and, therefore, in oscillation studies one
should better use the line-center intensities rather than the integrated
emission, which would allow intensity variations to mix waves from different depths.
More details on (i) the chromospheric oscillations in bright network and internetwork
regions, (ii) the phase shift and wave properties, and (iii) the heating of the
higher chromosphere at the sites of the network and internetwork regions, are
not discussed here, since the theme of the present paper is the
detection of mesocells at the upper chromosphere.

\subsection {Detection of mesogranulation}

We have derived the spatial power spectrum
plots along the slit for a given time of observation.
For example, in Figures 2 and 3, we have shown the spatial power spectrum
plots for all the Lyman lines and continuum for the time intervals of around
2000 s and 4000 s, respectively.  The spatial power spectrum plots show
two significant power peaks.
 The first peak is around 9-10 arcsec, which is in
consistent with the scale  of mesogranulation observed in the photosphere from SOUP observations
(November et al. 1981), in the temperature minimum region from TRC ultraviolet
 filtergrams (Martic et al. 2000, in preparation),
 as well as in the lower
chromosphere from \ion[Ca ii]  H \& K lines (Dam\'{e} \& Martic 1987, 1988,
Kariyappa et al. 1994),
and the second peak is seen at 4 arcsec, which may be related to
the width of the supergranular boundary.  In addition to these two peaks,
we find the smaller amplitude peaks at higher spatial periods.  They may be related
to the sizes of the supergranular cells.
We have looked for potential instrumental effects that could create periodic
signals in the spatial domain, and have analyzed the detector non-uniform
response, often referred to as flatfield.  The micro-channel-plate of the
detector has more or less regular geometrical structures which lead to
variations of the effective size of the pixels.  The corresponding effect on
the detector responsivity, which is been taken care of by the flatfield matrix revealed
a clear peak near pixel 26, but no prominent peak at higher frequencies.  This
excludes that residual effects of the flatfield correction could possibly be
responsible for the signals we have revealed and demonstrates their solar nature.
Therefore, we conclude that the maximum
around 9-10 arcsec corresponds to mesogranulation.
This result from SUMER observations of higher hydrogen Lyman lines is new
and suggests a mesogranulation structure in the upper chromosphere.
After carefully examining the power peak values,
it is interesting to mention that the size of the mesocells
will be the same in all the Lyman
lines including continuum, which are formed at different depths in the
chromosphere.
 We can remark that further up in the atmosphere this 9-10 arcsec dimension
will keep the same spatial extent and even up to the transition zone (Dere, Bartoe, \& Brueckner
1986), where only a slight increase (11 arcsec) in dimension.
The mesogranulation structure has been seen all along in the photosphere,
temperature minimum region,
in the lower and upper chromosphere, and this result suggests that there may be an
overshooting of
convective motions from layers  below.

\begin{figure}
   \centering
  \includegraphics[width=13.0cm,height=9.5cm]{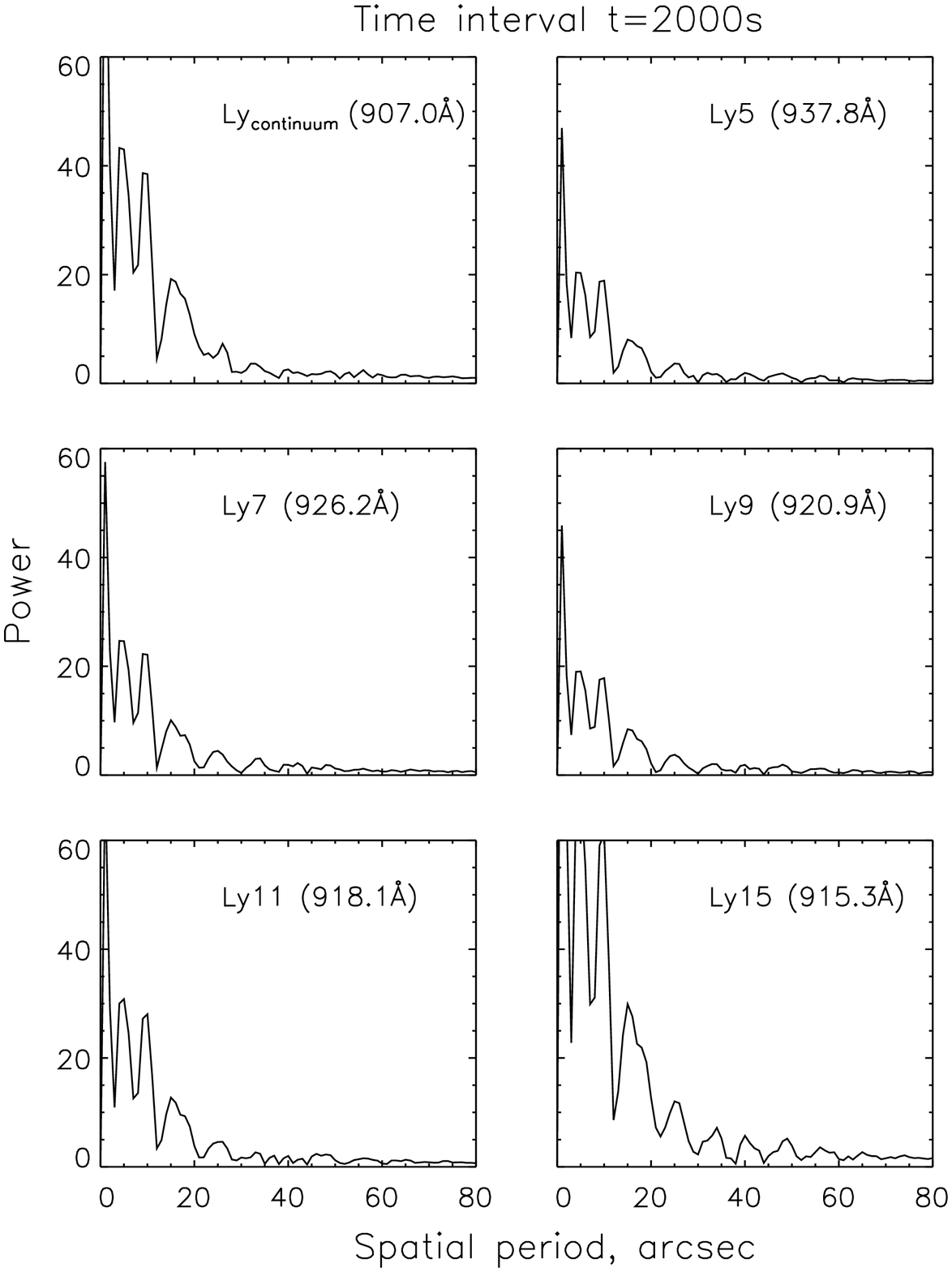}
   \caption{Same as Fig.2, but the spatial power spectra for the time
interval around 4000 s during our observing sequence.  Note that the scale and amplitude
of the peaks are more or less the same as in the case of 2000 s.  This suggests that the
mesocells
will have life time more than our total observing period.}
\label{FigVibStab}
    \end{figure}

        The observed
periodical brightenings on a meso-scale spatial extent above the granulation
certainly gives us an additional information on the processes governing the
structures in that region of the solar atmosphere.  The relationship between the
time evolution and periodical brightenings
of the mesogranulation has to be confirmed with still a longer duration of
observations.
 We have checked the spatial power spectrum taken for all the lines
randomly in time for
the entire duration of our observations.
We find that the size of the mesocells will appears to be
the same in all the spectral lines and at the continuum.
The amplitude
of the power peak will remain the same for a given line and time during our
observations.  Except that we will see a slightly larger amplitude in the power
peaks in the case of Lyman continuum ( at 907.0 \AA) and Ly15 (915.3 \AA)
line compared to the rest of other lines, and these two are formed rather
in the lower atmosphere.  From this result, we can infer that the life time of the
mesogranules will be more than several hours or more.
This is in good agreement with the findings reported by
Muller et al. (1992).  However, in order to check: (i) Whether we could see another
mesocell
at the previous location or not; (ii) Whether it is changing with time and
different positions over the solar disk, we need to investigate
further with the new observations.  The related research is in progress.
 Since the life time of the mesocells seems to be more than 3 hours,
our present time sequence of
observation is too short to comment on the temporal evolution of the
mesogranulation structures.

\begin{acknowledgements}
The SUMER project is financially supported by DLR, CNES, NASA, and the
ESA PRODEX Programme (Swiss Contribution).  SUMER is part of {\it SOHO}, the {\it SOlar
and Heliospheric Observatory}, of ESA and NASA.  We highly appreciate Dr. Luc
Dam\'{e}'s stimulating discussion on this research.  We wish to thank
Dr. Klaus Wilhelm for critical comments and suggestions, which helped
to improve the manuscript.
\end{acknowledgements}

\end{document}